\documentclass[twocolumn,showpacs,amsmath,amssymb,aps,prd]{revtex4}
\usepackage{graphicx}
\begin{document}
\title{Transition from inspiral to plunge for eccentric equatorial Kerr orbits}
\author{R. O'Shaughnessy}
\email{oshaughn@caltech.edu}
\affiliation{Theoretical Astrophysics, California Institute of Technology, Pasadena, CA 91125}
\date{Received ?? Month 2002}
\begin{abstract}
Ori and Thorne
have discussed the duration and observability (with LISA) of the transition from circular,
equatorial inspiral to plunge for stellar-mass objects into supermassive
($10^{5}-10^{8}M_{\odot }$) Kerr black holes. We extend their computation to eccentric Kerr
equatorial orbits. Even with orbital parameters near-exactly determined, we find that there is
no universal length for the transition; rather, the length of the transition depends
sensitively --- essentially randomly --- on initial conditions.  Still, Ori and Thorne's
zero-eccentricity results are essentially an upper bound on the length of eccentric transitions
involving similar bodies (e.g., $a$ fixed).  Hence the implications for observations are no
better:  if
the massive body is $M=10^{6}M_{\odot}$, the captured body has mass $m$, and the process occurs
at distance $d$ from LISA, then $S/N \lesssim (m/10 M_{\odot})(1\text{Gpc}/d)\times O(1)$, with the precise
constant depending on the black hole spin. For low-mass bodies ($m \lesssim 7 M_\odot$) for which the
event rate is at least vaguely understood, we expect little chance (probably [much] less than $10\%$,
depending strongly on the astrophysical assumptions) 
of LISA detecting a transition event with $S/N>5$ during its run;
however, even a small infusion of higher-mass bodies or a slight improvement in LISA's noise
curve  could potentially produce $S/N>5$ transition events during
LISA's lifetime. 
\end{abstract}
\maketitle

\section{Introduction}
The gravitational waves emitted during inspiral and infall of a body (mass $m$) 
into a black hole (mass $M$) should reveal detailed information about
the orbital geometry and the hole's  spacetime geometry,
 thereby providing high-precision tests of general
relativity \cite{Fintan}. 
While the scattering of waves off background curvature implies that waves emitted at any time during
the inspiral provide some small measure of even the smallest scale variations in the background
 spacetime geometry, the waves emitted as a particle passes through a region provide the most
sensitive tests of that region: they reveal what path the particle has followed, and therefore
constrain the spacetime to permit such a path.  Therefore, to provide a sensitive probe of the
innermost regions of black-hole spacetimes, we want to study orbits that pass as near as
possible to the hole itself.
Unfortunately, this means that the signals that are potentially the most 
informative about the hole's innermost structure are typically  the briefest: they arise from the
end of the bound portion of the orbit and from the transition from
inspiral to plunge. 
Since the relevant fraction of the orbit  persists for only a small  fraction of the overall
detectable inspiral, we have significantly less probability to resolve waves during this
interval than to resolve earlier, longer portions of the inspiral.  One therefore wants to roughly characterize
the waves emitted during these intervals (in the case of LISA sources, the goal of this paper).
If this  characterization suggests that 
planned observatories such as LIGO or LISA could detect them, one should then carry out much more detailed
studies of these last few orbits and the waves they emit.

For inspirals appropriate to the LIGO band ($\sim 10$-$10^3$ Hz) and which LIGO can
plausibly detect ($\eta =m/M$ $\sim 0.01$ to 1),  order-of-magnitude
computations (say, by post-Newtonian methods) suggest the last few waves are 
detectable \cite{ParametricFit}.  
But because in this regime simple approximation techniques (such as
post-Newtonian \cite{ParametricFit,Alessandra,KWW} or
test-particle approximations) break down, and because numerical
relativity \cite{num1} codes
remain incapable of evolving orbits accurately enough to find the waves, the community does not
yet possess a waveform trusted for any purpose beyond detection.

LISA's band ($\sim 10^{-3}$-$10^{-1}$Hz)  will prove more sensitive to
extreme mass-ratio infalls --- that is, to stellar-mass black holes, white dwarfs, and neutron
stars falling into supermassive [$M=O(10^{5}-10^{8})$, so $\eta \sim 10^{-4}$ to $10^{-8}$] black holes \cite{LISA}. With such extreme mass
ratios, the computation of detailed waveforms for purposes beyond mere detection 
should prove much simpler:
to understand evolution, we need do nothing more
than solve the classical radiation-reaction problem, albeit on a curved spacetime and with a
gravitational, rather than electromagnetic, field \cite{reaction}.
While this problem hasn't been solved to the accuracy required to
construct long-integration-time coherent detection templates,
 one
can employ adiabatic approximations to address most preliminary investigations. For example, 
as Ori and Thorne \cite{OriThorne}
have discussed in the context of circular inspiral, to understand the $\eta \ll 1$ transition's
duration --- measured in experimentally observable gravitational wave cycles --- we do not need a
precise knowledge of the reaction force. An averaged reaction force ---- one we can easily
deduce from the radiation of conserved constants ---- suffices for the short interval we will
coherently employ it.
Applying this reaction force, we can follow the particle through transition and thereby
predict roughly how long this transition will last. 

The goal of this paper is to  extend the Ori-Thorne analysis
to eccentric Kerr orbits,  in an effort to estimate the prospects of LISA detecting a
transition from inspiral to plunge.

This analysis relies on using the radiation of two conserved constants $E,L$ to compute the
effect of radiation on the orbit. But for Kerr inclined orbits there is an additional constant ---
the Carter constant --- whose evolution has not yet been  related to fluxes
at infinity.  
Since we 
lack the necessary tools, we leave the Kerr inclined case to a future paper.

\subsection{Outline of this paper and summary of conclusions}
In Sec. \ref{sec:framework}, we will outline the basic physical framework behind our
approach. In particular, we will introduce an explicit procedure to estimate the time duration of a transition.
This procedure takes as input the net (time-averaged) fluxes of energy and angular momentum
from the particle's instantaneously geodesic orbits, input one obtains from 
a solution of the Teukolsky equation given  a geodesic orbit as source.
This procedure also takes as
input some observationally-defined interpretation of what ``the transition region'' is.  As the
latter is ambiguous, and depends on exactly what sorts of templates one uses to find it, the
exact length of the transition will depend on the convention one uses.

Ideally, one should define some unambiguous set of templates and match those against the
simulated emitted waves to both define the transition duration and deduce the resulting
signal-to-noise ratio for a given source. But for brevity and simplicity, as discussed
in Sec. \ref{sec:framework2}, we will use a much cruder scheme --- based on a purely
sinusoidal, quadrupolar model for the waves ---  to characterize the expected LISA signal-to-noise ratio
from a specific transition crossing.  Given $S/N$ for an event and loosely-understood rates for
transition events, we then develop, in Sec. \ref{sec:framework2}, a scheme for estimating the probability that LISA will see an event with $S/N$
greater than some detection threshold.
 
With this complete scheme for  estimating the signal-to-noise of a characteristic source 
 and  determining the probability that LISA, in its currently-planned configuration, will
see something, in Sec. \ref{sec:sch} and Sec. \ref{sec:kerr} we will apply it to inspirals into
Schwarzchild and Kerr holes, respectively.
We find in Sec. \ref{sec:detectFixedSource} that  Ori and Thorne's
zero-eccentricity results are essentially an upper bound on the length of eccentric transitions
involving similar bodies (e.g., $a$ fixed).  It follows, in
Sec. \ref{sec:detectSomething},  that if we accept current (rough) astrophysical estimates of
the masses and numbers of inspiralling stellar-mass black holes and if we employ only the current LISA design, we
expect LISA will not see any transitions from inspiral to plunge during its lifetime, though it
may come close.

Slight changes in LISA could make some transitions detectable.
Dramatic improvements would be required to render LISA sensitive to prograde inspirals of
stellar-mass black holes into rapidly-spinning ($a>0.9$) supermassive holes.
But assuming such inspirals are a small proportion of all inspirals,
if the LISA noise curve
is lowered by a factor 3 (as is under currently discussion for other reasons), or
if nature provides black holes more massive than $10 M_\odot$ (say $30
M_\odot$) in numbers approaching current estimates for $10 M_\odot$,  LISA would have a good
chance of seeing one or two transitions sometime during  its lifetime. 

\section{\label{sec:framework}Physical framework underlying the transition length estimate}
In the (formal) absence of radiation reaction, a particle in equatorial
orbit about a Kerr hole moves along a geodesic. Its
radial motion can be determined from a first integral of the geodesic
equation (equivalent to
conservation of rest mass; see comments in Appendix \ref{ap:evolveMax}) 
\cite{MTW}:
\begin{equation}
 \left( \frac{dr}{d\tau }\right) ^{2}+V\left[ r\left(\tau \right) ,E,L\right]  =0 ,
\label{eq:geodesic}
\end{equation}
\begin{equation}
V\equiv-\left( E^{2}-1\right) -\frac{2}{r}
+\frac{\left(L^{2}-a^{2}\left( E^{2}-1\right) \right)}{r^{2}}
-\frac{2\left( L-aE\right) ^{2}}{r^{3}}  .
\label{eq:potential}
\end{equation}
Here and throughout this paper all all quantities are, for simplicity, made  dimensionless
using the particle's mass $m$ and the hole's mass $M$:
$E=$(orbital energy)/$m$, 
$L=$(orbital angular momentum)/$mM$, $r=$(orbital boyer-lindquist radius)/$M$, $\tau=$
(particle's proper time)/$M$, and $a=$ (hole spin angular momentum)/$M^2$. 
Physical solutions may be specified by ($E$,$L$) or by any
other pair of equivalent orbital parameters. 
It is conventional in the inspiral literature to employ as alternatives the parameters $p$ (a relativistic
generalization of semi-latus rectum) and $e$ (a relativistic generalization of orbital eccentricity)
\cite{CKP,Dan}; these parameters are discussed in more detail in
  Appendix
\ref{ap:kerrbasics}. 

\begin{figure}
\includegraphics{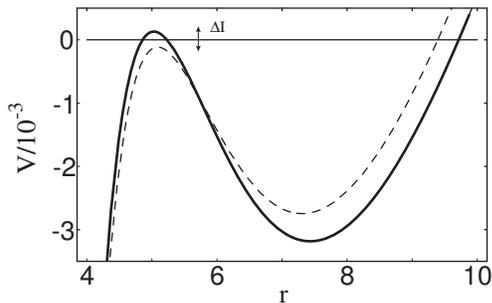}
\caption{\label{fig:criticalcurves}The effective potential $V(r,E,L)$ for radial geodesic motion
gradually evolves during inspiral. If the initial eccentricity is nonzero, the maximum of the
 potential will gradually decrease until it passes below zero, thus permitting the particle to
 fall into the hole.  By way of illustration, we plot $V(r,E,L)$ for $a=0$ and
 (solid) ($E$,$L$)=(0.948157, 3.53038) and (dashed) ($E$,$L$) =(0.947454, 3.52092). Under
 the action of radiation reaction, if $\eta=2\times 10^{-3}$ (an exaggerated mass ratio)
 the
 first system will evolve into the second after one radial period (of the first potential).
}
\end{figure}

We concern ourselves with a region of parameter space for which the maximum $V_\text{max}$ of the potential is
nearly 0 (Fig. \ref{fig:criticalcurves}) and which therefore nearly admits a circular orbit at
the radius $r_\text{max}$ of the maximum.  The geodesic equation Eq. (\ref{eq:geodesic}) implies that
particles can spend an extremely (logarithmically) long time near the maximum; i.e. 
the particle can ``whirl'' several times about the hole in angle without moving
significantly in $r$.
It is conventional to call this portion of the orbit the ``whirl.''

In the presence of radiation reaction (henceforth assumed weak), we must add to the geodesic
equation (gauge-dependent) time-varying terms which reflect the (gauge-dependent) influence of
gravitational radiation on the test particle's path.  These gauge-dependent terms oscillate on
the same characteristic timescale as the radiation field.
Since the radiation field is predominantly produced during the whirl part of the orbit,
the radiation field predominantly oscillates at harmonics of the angular frequency
$\Omega$ of circular orbits at the maximum.
By averaging these reaction forces over a few cycles (e.g., over times $\approx 2\pi/\Omega$)
to obtain their secular effect, we in principle find expressions for $E(\tau)$ and $L(\tau)$.
Since the averaging time can still (particularly when the particle whirls several times about
the hole) be shorter
than the time the particle spends whirling around the central hole, we can to a good
approximation employ Eq. (\ref{eq:geodesic}) with \emph{time-varying} $E(\tau)$, $L(\tau)$ to
follow the orbit when the particle is near the maximum of the potential -- 
 and in particular the interval in
which the particle goes  
from nearly-geodesic bound orbit into rapid plunge into the hole.

In this paper, we do not compute $E(\tau)$ and $L(\tau)$ in the thorough, general manner
described here. See
Sec. \ref{sec:gettingInputs} for a discussion of what information
about $E(t)$ and $L(t)$ is required for our estimate and how that
information is obtained.

\subsection{\label{sec:sub:survey}Why we may still approximate the potential as static when computing the radial orbit}
According to Eq. (\ref{eq:geodesic}), a test particle on an approach to a black hole will fall
into the hole if the local maximum of $V$, 
\begin{equation} 
\label{eq:defineI}
I \equiv-V(r_{\max}) = - V_\text{max} \; ,
\end{equation}
is negative.
Radiation reaction  reduces the local maximum faster than
this potential $V$ flattens out.  As the maximum decreases, the particle spends ever-more of
its radial cycle near the local maximum. Eventually, we reach configurations such as those
shown in Fig. \ref{fig:criticalcurves}, where the particle can slip over the maximum and fall
into the black hole.

While configurations with $I\approx 0$ appear delicately balanced and therefore highly
sensitive to small changes in $(E, L$), in fact under weak conditions (conditions made more
explicit in Sec. \ref{sec:nonadiabatic}) one may ignore radiation reaction when computing a
radial orbit and treat the potential $V$ as static.

Suppose a particle starts its whirl with some values for $(E,L)$ (and therefore
$I$). During the whirl, even though $(E,L)$ change, the peak of the potential ($I$) will
not change significantly (see Appendix \ref{ap:evolveMax}). Moreover, the location of the peak
usually moves slowly relative to the particle.  A condition for when the latter holds is
presented in Sec. \ref{sec:nonadiabatic}. Therefore, during the whirl 
one can ignore radiation reaction.

After the particle finishes whirling about the hole, it moves outward to its outer radial
turning point and back. During this period, the maximum \emph{does} change, from $I$ to
$I' = I+\Delta I$. Unless the potential is nearly flat, however, the potential away from the
neighborhood of the hole will not change much as $(E,L)$ change.  Again,
a condition for when the latter holds is presented in Sec. \ref{sec:nonadiabatic}. 
Therefore, in this interval
one can again ignore radiation reaction.

When we attempt to evolve the particle through the next ``whirl'', we need the correct value of
the height of the maximum (now $I' =I+\Delta I$) to determine how long the particle whirls around the
hole.  Therefore, when we start the cycle anew, we must use a potential with parameters $(E+\Delta
E, L+\Delta L)$, with $\Delta E$ and $\Delta L$ the  change
in these  constants over the preceeding full radial period.
If $I'<0$, the particle will ``bounce'' off the maximum and we repeat the cycle above once more. 
But eventually we will have $I'>0$,
 at which point the particle will move across
the maximum during its whirl and will subsequently  ``plunge'' into the hole.

To summarize: so  long as the potential is approximately static
(cf. Sec. \ref{sec:nonadiabatic}), we expect we can understand transitions 
from inspiral to plunge by way of examining the geodesic equation [Eq. (\ref{eq:geodesic})] in
the neighborhood of the local maximum, using $I\in [0, \Delta I]$.

\subsection{\label{sec:sub:adiabatic}Adiabatic approach to estimating the duration of the transition from inspiral to plunge}
So long as we can treat the potential as static, we can approximate Eq. (\ref{eq:geodesic}) in
the neighborhood of the potential's maximum at $r=r_\text{max}$ by the form
\begin{equation}
\label{eq:geodesicQuadraticAdiabaticApprox}
 \gamma^2 \left(\frac{d \delta r}{dt}\right)^2 + \frac{\delta r^2}{\tau_o^2} = I.
\end{equation}
Here $\delta r \equiv r-r_{\max}$;  $r_{\max}$ is the instantaneously static location of the local
 maximum of $V$, and also the point about which we have expanded the potential;
\begin{equation}
\tau_{0}\equiv(V''/2)^{-1/2}
\end{equation}
is a constant related to the curvature of the potential at the transition location; $t$ is the
(dimensionless) time at infinity; and $\gamma$ is the redshift factor relating proper time $\tau$
to Boyer-lindquist coordinate time at $r=r_{\max}$:
\begin{equation}
\gamma = dt/d\tau{}_{r=r_\text{max}} \; .
\end{equation}

\emph{Estimating the duration of a given transition}:
Solutions to Eq. (\ref{eq:geodesicQuadraticAdiabaticApprox} give hyperbolic motion; for example, the solution appropriate to  $I>0$ is
\begin{equation}
\label{eq:basicHyperbolicCrossing}
r(t)-r_{\max}\approx \sqrt{I}\tau_{0} \sinh t/(\tau_{0} \gamma) \; .
\end{equation}
Using this solution, we conclude that the transition time going from $r-r_{\max}=-\delta
r_\text{ref}$ to $r-r_{\max}=\delta r_\text{ref}$ is
\begin{eqnarray}
\label{eq:generalCrossingTime}
T_{c}(\delta r_\text{ref},I) &\approx& 2 \gamma \tau_o \sinh^{-1} \left[ \delta r_\text{ref} / (\tau_o
 \sqrt{I}) \right ] \\ &\approx& 2 \gamma \tau_o \ln \left[ 2\delta r_\text{ref}/(\tau_o
 \sqrt{I}) \right ] \nonumber
\end{eqnarray}
Hence given $\delta r_\text{ref}$, a quantity which defines what we mean by ``the transition
extent'', we can estimate the length of any  transition (characterized by $I$) at any
 transition location (characterized by the explicit values that go into $\gamma$,
$\tau_o$).

\emph{Estimating the distribution of transition durations}: There is 
no unique transition duration. Rather, we have a distribution of durations,
depending on the distribution of $I$ at the start of the particle's final whirl.
But that distribution is simple:
  since an initial configuration of particles will have some distribution of
  $I$, since this distribution evolves smoothly with no ``knowledge'' of the preferred scale
  $\Delta I$, and since  $\Delta I$ will be smaller than any scale in the 
  distribution function, 
  a test particle on its final, plunge-triggering whirl has an approximately equal probability to have any 
  $I\in[0,\Delta I]$.  
Therefore,  the probability density for a test particle to have
  a given duration between $T_c$ and $T_c+ dT_c$ is $dP \propto dT_c (dI/dT_c) \propto dT_c
  \exp[-T_c/\gamma \tau_o]$; see Eq. (\ref{eq:generalCrossingTime}).
Denoting by 
\begin{equation}
\label{eq:mainGeneral}
T_{c-} = T_c(\delta r_\text{ref}, \Delta I) 
\end{equation}
the minimum possible transition duration, and ignoring the tiny regime of transitions which are
nonadiabatic (see Sec. \ref{sec:nonadiabatic} below), we conclude that 
\begin{equation}
\label{eq:mainGeneralDistribution}
dP \approx\Theta(T_c-T_{c-}) e^{-(T_c-T_{c-})/\gamma \tau_o} dT_c/\gamma \tau_o \; .
\end{equation}
[where $\Theta(x)=1$ when $x>0$, 0 otherwise].  

We can also characterize
distribution of crossing times by a function $T_c(p)$ such that only a fraction $p$ of particles
could (assuming the conditions of Sec. \ref{sec:nonadiabatic} hold) have longer crossing times.
For example, only a 
fraction $10^{-n}$ of particles will have duration longer than
\begin{equation}
\label{eq:mainGeneralCutoffs}
T_{c,n} \equiv T_c(\delta r_\text{ref}, \Delta I\; 10^{-n}) \approx 2 \gamma \tau_o \ln \left[
   \frac{2\delta r_\text{ref}}{\tau_o 10^{-n/2} \sqrt{\Delta I} } \right ] \; .
\end{equation}

\emph{Additional comments}: 
\begin{itemize}
\item \textbf{Converting to number of cycles:}
As the particle passes through the transition region, the particle ``whirls'' about the black hole a
few times. Since its radial location is largely fixed while it whirls around the hole, so is its angular frequency
$d\phi/dt\equiv \Omega$; therefore, 
we can re-express any duration $T_c$ in terms of a ``number of orbital cycles'' the particle
``whirls'' around the hole $N_c$, defined
by 
\begin{equation}
\label{eq:defineNc}
N_c = \frac{T_c \Omega}{2\pi}.
\end{equation}
Since we concern ourselves with only Kerr equatorial orbits, we have
\begin{equation}
\label{eq:defineOmega}
\Omega(r) = \frac{\text{sign}(a)}{r^{3/2}+a}.
\end{equation}

\item \textbf{Characteristic duration and variation of $T_c$ with $e$:}
  By examining the quadratic approximation to the potential
  [Eq. (\ref{eq:geodesicQuadraticAdiabaticApprox})], or equally well from Eq. (\ref{eq:generalCrossingTime}),
  we see
  that the transition duration is always $T_{c} \sim $ (few)$\times \gamma \tau_o$ ---
  that is, the
  crossing time is around the natural timescale of the effective potential.  Admittedly, since
  $O(\Delta I)=O(\eta)$, the quantity labeled (few) could be --- and will be --- significant; 
  therefore, the logarithmic correction in Eq. (\ref{eq:generalCrossingTime}) is necessary.  But for purposes of understanding the
  variation of crossing time with orbital parameters, largely we can regard
  $T_c \sim \gamma \tau_o$.
    For example, we expect $T_c$  to increase monotonically with decreasing orbital eccentricity $e$ ---
that is, as the maximum possible energy barrier decreases and the potential flattens out ---
simply because $\tau_o$ does.  [By way of example, see Eq. (\ref{eq:schGammaTau0}), an expression
for $\gamma \tau_0$ appropriate to Schwarzchild.]

\item \textbf{On variation of $T_c$ with $\eta$:}
Similarly, we can loosely characterize the dependence of the duration distribution --- or, for
clarity, $T_{c-}$ --- 
on $\eta$ by noting
i) $\sinh^{-1}(x)\approx \ln 2 x$ when $x$ is large and 
ii) $\Delta I \propto \eta$, so we can characterize variation with $\eta$ by $H(\eta_o)$, defined by
\begin{eqnarray}
\label{eq:mainVaryWithEta}
\frac{T_{c-}(\eta)}{T_{c-}(\eta_0)} - 1
    &\approx&
    \frac{\ln \sqrt{\eta_o/\eta}}{\ln \left( \delta r_\text{ref} / \tau_o \sqrt{\eta_o \Delta I/\eta }
    \right) } \nonumber \\
 &\equiv& \ln(\eta_o/\eta) H(\eta_o) \; 
\end{eqnarray}
[where we have used the fact that $\Delta I/\eta$ is independent of $\eta$ to justify writing
the denominator as $2/H(\eta_o)$].
In other words, while the minimum transition duration will grow slightly shorter with larger mass
ratios, the dependence (like the dependence on $\Delta I/\eta$) is weak; typically (e.g., for
Schwarzchild) 
we find $H\in \sim [0.1, 0.4]$.

\end{itemize}

\subsection{\label{sec:nonadiabatic}Explicit conditions under which we may continue to
  approximate the potential as static}
Throughout our analysis, we have approximated the potential as static. As outlined in
Sec. \ref{sec:sub:survey}, there are two ways in which this approximation could fail.

First, the potential away from the maximum could change significantly during one whole radial
orbit. Generally the change of $V$ at any specific location is small. Such changes therefore
matter only if the potential is delicately balanced near
zero at every point in which the particle orbits. More explicitly, we expect problems if the change
$\Delta I$ of the potential's maximum during one whole radial orbit is comparable to the difference between the maximum and minimum of $V$. Therefore, we
conservatively require 
\begin{equation}
\label{eq:defineImax}
 I_\text{max} \equiv V(r_\text{max}) - V(r_\text{min}) \gg \Delta I \; . 
\end{equation}
An explicit form for $I_\text{max}$ is presented in Eq. (\ref{eq:generalImax}).
Since the potential gets very flat as $e\rightarrow 0$, our approximations will break down at eccentricities below $e_\text{min}$, defined by solutions to
\begin{equation}
\label{eq:defineEad}
\Delta I = I_\text{max}(e_\text{min}) \; .
\end{equation}

Second, the radial location $r_\text{max}$ of the maximum could move significantly while the particle is
in its last whirl about the hole. Based on  Eq. (\ref{eq:geodesic}), to prevent against this we
require $I=(dr/d\tau)^2 \gg (dr_\text{max}/d\tau)^2$, i.e. that
  \begin{equation}
    \label{eq:defineIadmin}
   I\gg I_\text{ad,min} \equiv \left(\frac{d r_\text{max}}{d\tau}\right)^2  
    =  \left(\gamma \frac{d r_\text{max}}{d t}\right)^2  \; .
  \end{equation}
The precise  procedure that we will use to estimate $d r_\text{max}/d\tau$ will be
discussed in Sec. \ref{sec:gettingInputs}.
In summary, so long as $I\gg I_\text{ad,min}$,
 or equivalently so long as the crossing duration significantly shorter
than 
\begin{equation}
\label{eq:defineTad}
T_{c,\text{ad,min}} = T_{c}(\delta r_\text{ref}, I_\text{ad,min}) \; ,
\end{equation}
 gradual motion of the  potential will not
significantly alter the transition length estimates presented earlier.

\subsection{\label{sec:gettingInputs}Inputs necessary for estimating the 
  transition length}
In the above we have outlined a computational procedure which takes as input $\delta
r_\text{ref}$ and knowledge about radiation reaction (namely, about $\Delta I$ and about $dr_\text{max}/dt$)
 and which gives us in return an estimate
of the length of any specific transition from inspiral to plunge.
We now describe the explicit approximations we shall
use to estimate $\Delta I$ and $I_\text{ad,min}$ from known information about $E(\tau)$ and
$L(\tau)$. We also make an explicit choice for $\delta r_\text{ref}$.

\subsubsection{\label{sec:estimateDI}Estimating \protect{$\Delta I$}}
As described in Sec. \ref{sec:sub:survey}, we obtain $\Delta I$ by comparing the potential  $V$
when the conserved constants are $(E,L)$ to the potential $V$ when they are $(E+\Delta E,L+\Delta L)$, where $\Delta E$ and
$\Delta L$ are the change in the appropriate conserved constants over one radial orbit.
We obtain $\Delta E$ and $\Delta L$ from numerical solutions to the Teukolsky equation. From their code,
Glampedakis   and Kennefick have kindly provided time-averaged fluxes $\left<dE/dt\right>$
and $\left<dL/dt\right>$  \cite{Dan}, which, when  combined with an expression for the radial
period $T(E,L)$ as given in any  classical relativity text pe.g., Eq. (33.37) of MTW
\cite{MTW}], yields  $\Delta E$ and  $\Delta L$, and thus $\Delta I$ [Eqs. (\ref{eq:potential})
and (\ref{eq:defineI})].
In this fashion, for each black hole (parametrized by spin parameter $a$), we can find $\Delta
I(p,e,a)$ for any 
equitorial geodesic with parameters $(p,e)$. 

\begin{figure}
\includegraphics{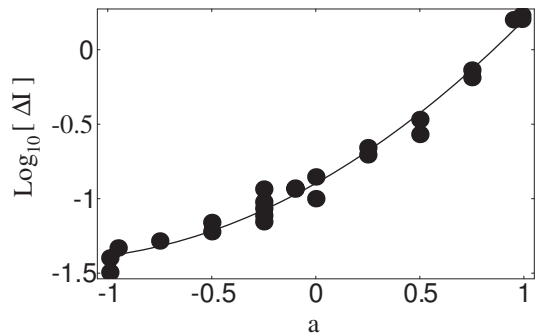}
\caption{\label{fig:deltaIcurve}The (log of the) change of the potential's maximum during the
  last radial orbit (\protect{$\log_{10}[\Delta I/\eta]$})
  versus $a$.  
The points show our  $\Delta I$ for various cases with parameters $(p,e)$ close to
those consistent with circular orbits (the boundary between stable and unstable orbits); these
  points were obtained from numerical solutions of the Teukolsky equation by Glampedakis and
  Kennefic, using the
procedure described in Sec. \ref{sec:estimateDI}. For each $a$, solid circles show values of 
\protect{$\log_{10}[\Delta I/\eta]$} for several different values of $e$; that these points are
  all consistent with a single fit demonstrates that $\Delta I$is approximately independent of
  $e$.
The solid curve is the quadratic fit Eq. (\ref{eq:defineDeltaIFit}).
[In cases where insufficient data was present for extrapolation to the line $p=p_s(e,a)$, solid
  circles also indicate rough upper and lower limits expected of \protect{$\log_{10}[\Delta I/\eta]$}.]
}
\end{figure}

We need  $\Delta I$ only for the particle's last
whirl. Since the maximum is extremely close to zero, the orbital parameters $(p,e,a)$ nearly satisfy a
condition for the existence of (unstable) circular orbits $p=p_s(e,a)$ (see Appendix
\ref{ap:kerrbasics}).  This curve also necessarily serves as the boundary between stable orbits
and plunge.
In the vicinity of this boundary line, 
 $\Delta I$ is well-approximated by its nonzero values on the boundary.  So for our
 computation we seek an
 expression $\Delta I(e,a)$.

In practice, from the values of $\Delta I$ at points near this boundary line, we extrapolate to estimate
$\Delta I$ on the boundary surface itself.
Figure \ref{fig:deltaIcurve} shows the results of our extrapolation.

One can argue that
$\Delta I(p,e,a)$ on the last-stable-orbit boundary $p=p_s(e,a)$ should largely be
independent of $e$ at moderate eccentricity \footnote{Since the orbit is nearly circular,
  radiation of conserved constants should be nearly uniform in time, so assume $E\approx E_o +
  t\times dE/dt $ and similarly for $L$.
Take a third-order approximation to the
  potential.  Find an explicit expression for $dI/dt$ in terms of the solution $r(t)$ and the
  motion of the maximum $r_\text{max}$.  Use an approximate (sinusoidal+constant) solution
  for $r(t)$ in the previous expresion to show that $\Delta I$ over one radial period is
  approximately independent of eccentricity.}.
For this reason, Figure
\ref{fig:deltaIcurve} shows results only as a function of one parameter ($a$).
Numerical data over the range $e\in[0,0.5]$ support this conjecture.  
Therefore, so long as we avoid $e\approx 1$, where this conjecture
has not yet been tested and likely fails, we can approximate
 $\Delta I/\eta$ by a function independent of $e$.
Fitting a relatively simple function (exponential form in $a$, independent of $e$) to the data
in Fig. \ref{fig:deltaIcurve}
we find
\begin{equation}
\label{eq:defineDeltaIFit}
\log_{10}\frac{\Delta I}{\eta} = -0.8972+0.7911 a + 0.3047 a^2 \; .
\end{equation}

\subsubsection{Estimating \protect{$I_\text{ad,min}$}}
To evaluate $I_\text{ad,min}$, we need no more than i) knowledge of the potential (which tells us
 $r_\text{max}$ as a function of $E$, $L$) and ii) knowledge of $dE/dt$, $dL/dt$
when the particles are in nearly-circular orbit near the hole. 

In principle, we could approximate the latter
by the appropriate values for an exactly  circular (unstable) orbit. 
As a practical matter, 
comprehensive tabulation of
the physically appropriate instantaneous
$dE/dt$ and $dL/dt$ for all transitions of interest --- namely, the values appropriate to a
circular unstable orbit --- proves time-consuming and technically challenging.   
Furthermore, because the crossing time depends only weakly (logarithmically) on $I$, and because
exceedingly few particles will have $I\lesssim I_\text{ad,min}$,
we only need $I_\text{ad,min}$ to order of magnitude.

Therefore, for
practical purposes, when estimating $I_\text{ad,min}$ by way of Eq. (\ref{eq:defineIadmin}) we will i) perform the computation for $d
r_\text{max}/d\tau$ analytically in terms of $dE/dt$ and $dL/dt$, ii) simplify under the
assumption  $dE=\Omega dL$, which would be valid if we used the true forms for $dE/dt$ and
$dL/dt$,  and then iii) 
insert for $dL/dt$ the Peters-Mathews expression (an estimate obtained using linearized,
quadrupolar emission from newtonian orbits) \cite{PetersMathewsRadiation}
\begin{equation}
\label{eq:dLdtPeters}
\frac{dL}{dt} \approx \eta \frac{32}{5}\frac{1}{p^{7/2}} (1-e^2)^{3/2}
  \left[
    1+\frac{7}{8} e^2
  \right] \; .
\end{equation}

\subsubsection{Choosing \protect{$\delta r_\text{ref}$}}
To complete our procedure, we must define ``the'' transition duration.
Unfortunately, because ``the'' transition from inspiral to
plunge occurs at no definite location, has no well-defined start or 
finish,
the transition duration remains a matter of convention \footnote{The closest
``natural'' definition would be some fraction, defined some way or another, of the
length of the binding region. But since the binding region goes to zero length, when
the potential gets flat, the length of the transition would go to zero. 
We therefore would have the unusual result that the transition from circular inspiral to plunge
took no time. This result is inconsistent with the Ori \& Thorne value.}.
We shall adopt a convention motivated by a simple model of gravitational-wave data analysis.

The key feature of waves emitted during the transition is their considerable 
simplicity: they are emitted from a nearly-circular-equatorial orbit at $r_\text{max}$, and hence
are characterized by the angular frequency $\Omega$ associated with circular 
orbits there.  If we were to try to detect these gravitational waves --- for
simplicity, focusing on the dominant frequency component, $\omega = 2\Omega$ 
--- we would want to insure that our model $\phi = 2 \Omega t$ for the
gravitational wave phase agrees, within $\pi$, with the true wave phase.

The true rate of change of orbital phase is 
\begin{equation}
\label{eq:instantaneousFrequency}
\frac{d\phi}{dt}(r,E,L) = \frac{g^{\phi\phi} L - g^{\phi t} E}{-g^{tt} E + g^{t\phi}L} \; ,
\end{equation}
where $g^{\alpha\beta}$ are known Kerr metric functions in Boyer-Lindquist coordinates, and $E$, $L$
are consistent with the circular orbit at $r=r_\text{max}$ (use standard expressions for
$E$, $L$ appropriate to circular orbits, such as Eqs. (2.12) and (2.13) of Bardeen, Press, and Teukolsky \cite{CircularOrbits}).
Demanding that the difference between the true angular phase and our fiducial reference
$2 \Omega t$ be no more than  $\pi$ over  the length of the transition, we find a constraint on
the crossing duration $T_c$:
\begin{equation}
\pm 2\pi/4 = \int_{-T_{c}/2}^{T_{c}/2} \left[
   \frac{d\phi}{dt}(r(t)) - \frac{d\phi}{dt}(r_{\max})
   \right] dt.
\end{equation}
When we insert  $r(t)-r_{\max} = A \sinh \left[ t/\tau_o
  \gamma \right ]$  
with $A=\sqrt{I}\tau_o$ [Eq. (\ref{eq:basicHyperbolicCrossing})] into the above, we find an expression we can invert 
for $\delta r_\text{ref}(\Delta I)$:
\begin{equation}
\pi/4 \approx  \left|\frac{d (d\phi/dt)}{dr}\right| \gamma \tau_o \sqrt{ \delta r_\text{ref}^2 + A^2 } 
      \approx  \left|\frac{d (d\phi/dt)}{dr}\right| \gamma \tau_o \delta r_\text{ref} 
\end{equation}
[where the constant $A$ has been neglected in this expression, as it is always much smaller than $\delta r_\text{ref}$].
Solving for $\delta r_\text{ref}$, we obtain
\begin{equation}
\label{eq:drMonochromatic}
\delta r_\text{ref} = \frac{\pi}{4 \gamma \tau_o |d (d\phi/dt)/dr|}.
\end{equation}
We will use this form even when it predicts $\delta r_\text{ref}=O(1)$ [in other words, when
$\delta r_\text{ref}=O(M)$ when we convert to physical units].
Notice this $\delta r_\text{ref}$ is independent of mass ratio.

\section{\label{sec:framework2}Estimating the probability for  LISA to  observe a transition}
We wish to  estimate, for each choice of the supermassive hole's  angular
momentum and  distance from earth, and for each choice of test 
particle orbital parameters, the signal-to-noise ($S/N$) LISA would obtain  from waves emitted
during 
the  transition.  By combining this $S/N$ with the
 (poorly-known) statistics of
black-hole inspirals, we can estimate the probability LISA will see a transition event (e.g., have
$S/N > 5$).

\subsection{Estimating LISA's signal to noise for a given transition}
Since the transition waves  are  emitted by a circular orbit of frequency
\begin{equation}
\label{eq:defineForb}
f_\text{orb} = \Omega(r_\text{max})/2\pi M,
\end{equation}
the gravitational waves will be at that frequency and its
harmonics. For simplicity, assume that LISA detects only the strongest waves, the waves emitted
from the second harmonic $f_\text{tr}=2 f_\text{orb}$. 
These waves will last for an interval 
\begin{equation}
\label{eq:defineDT}
\Delta t = N_c/f_\text{orb}(=M T_c) \; .
\end{equation}
We can approximate their  characteristic rms
(source-orientation-averaged) amplitude 
 [following OT equation (4.7)] as a Peters-Mathews-style quadrapole term
(averaged over all orientations) times
a relativistic correction:
\begin{equation}
\label{eq:defineHrms}
h^\text{rms} = \frac{8}{\sqrt{5}} \frac{M}{d} \eta \Omega(r_\text{max})^{2/3} 
   \sqrt{\dot{\cal E}_{\infty, 2}} \; .
\end{equation}
Here $d$ is the distance to the source and $\dot{\cal E}_{\infty, 2}$ is a relativistic
correction factor defined explicitly in OT equation (2.3).

LISA has a spectral density of noise $S_h$ for waves incident on it with optimal propagation
direction and polarization; 
it has spectral density $5 S_h$ for typical directions and polarizations. Therefore, on
average, LISA should accumulate a signal-to-noise from the transition event given by 
\begin{equation}
\label{eq:detectformula}
\left( S/N\right) _{tr} = \frac{h^\text{rms}}{\sqrt{5 S_h(f_\text{tr})/\Delta t}}  \; .
\end{equation}

Particularly special  sources could have significantly higher $S/N$. For example, we
can pick up an increase of 
$\sqrt{5}$ if the source is ideally positioned on the sky, and a similar increase if the
source itself is optimally oriented.
But overall, the above scheme suffices to  estimate the signal-to-noise LISA would see from the
transition 
between inspiral and plunge for any capture $m$ into $M$ with any specific source
parameters (e.g., $e$, $a$) at any distance $d$.

\subsubsection*{Explicit expressions needed to compute LISA's signal to noise for a given
  transition}
To evaluate Eq. (\ref{eq:detectformula}), we need in addition to  $N_c$
and $\Omega$ [which enter into $S/N$ via  $\delta t$ and $f_\text{tr}$] the LISA noise curve $S_h$ and the relativistic correction factor
$\dot{\cal E}_{\infty, 2}$.
The LISA noise curve may be modeled by [OT equation (4.9)]
\begin{eqnarray}
\label{eq:lisanoiseOT}
  S_h(f)&=&\left[ 
    (4.6\times 10^{-21})^2 + (3.5\times 10^{-26})^2 \left(\frac{1\text{Hz}}{f}\right)^4
      \right. \nonumber \\
   & & \left. 
  + (3.5\times 10^{-19})^2 \left(\frac{f}{1\text{Hz}}\right)^2
 \right] \text{Hz}^{-1}\; .
\end{eqnarray}
The appropriate relativistic correction factor 
$\dot{\cal E}_{\infty, 2}$ can in principle be extracted from
simulations  of waves emitted by particles in unstable circular orbits. 
As in practice the latter proves 
time-consuming to evaluate and tabulate for all possible eccentric orbits and for all $a$, 
for simplicity we will assume that the appropriate relativistic correction factor is i) fixed
for all orbits close to a black hole of angular momentum  $a$ and ii) given explicitly by the
value appropriate to the innermost stable circular orbit (ISCO).
This latter expression has been tabulated by Ori and Thorne (see the $\dot{\cal E}_{\infty,2}$
column in their Table II); we approximate their
results by 
\begin{eqnarray}
\label{eq:otApproxE2}
\log_{10} \left( 
  \dot{\cal E}_{\infty, 2}
       \right)_\text{OT}  
   &\approx&   
   -0.0473+0.211x -0.053 x^2 \nonumber \\
  & & + 0.034 x^3 + 0.010 x^4
\end{eqnarray}
where $x=\log_{10} (1-a)$.

\subsubsection*{Dominant terms in the signal-to-noise estimate}
As written, the signal-to-noise estimate Eq. (\ref{eq:detectformula}) disguises what kinds of
effects predominantly influence it --- for example, whether changes in the strength of radiation
emitted prove more important or less than changes in the duration $T_c$ of the transition.
To clarify the dominant contributions to our estimate, fix some $a$ and compare the
signal-to-noise between two transitions ($1$,$2$) involving otherwise arbitrary parameters
(e.g., $m$, $M$, $d$, $e$).
Substituting expressions for $\Delta t$ [Eq. (\ref{eq:defineDT})], $h^\text{rms}$
[Eq. (\ref{eq:defineHrms})], and $f_\text{tr} = 2 f_\text{orb}$ [Eq. (\ref{eq:defineForb})] into 
Eq. (\ref{eq:detectformula}); assuming  
$\dot{\cal E}_{\infty, 2}$ is a fixed function of $a$; and comparing the resulting $S/N$ at two
sets of orbital parameters, we find 
\begin{eqnarray}
\frac{(S/N)_2}{(S/N)_1}& =& \frac{m_2}{m_1} \frac{d_1}{d_2}\sqrt{\frac{N_{c,2}}{N_{c,1}}} 
   \left( \frac{\Omega_1}{\Omega_2} \right)^{1/6}  \nonumber \\
\label{eq:detectformulaRelative}
& &  \times
 \sqrt{\frac{M_1 S_h\left(\frac{\Omega_2}{\pi M_2}\right) }{M_2
   S_h\left(\frac{\Omega_1}{\pi M_1} \right) }} .
\end{eqnarray}

The first two terms reflect the natural $m/d$ scaling of emitted waves. The third term
reflects the fact that more orbits around the hole during the transition mean more
gravitational wave cycles seen by LISA.  The fourth term, which combines the fact that
gravitational waves emitted closer to the hole are stronger and yet last for less time,
is to a good approximation constant.  Finally, the last term reflects LISA's
sensitivity.  The only term which depends explicitly on $M$ (ignoring the weak variation in $N_c$),
this last term selects black hole masses which have their transition  close to optimally positioned
in the 
LISA band, or $M\approx \text{(few)}\times 10^6$.  So long as the mass is so, this term varies
comparatively little.

\subsection{Method for estimating the probability of detecting some transition during  LISA's operation}
Above we gave a procedure for computing the $S/N$ for any given source.  
But the sources which
produce the strongest signals (inspirals very close by) are rare.
 Therefore, for any given $(S/N)_o$ we
have a certain probability that, during the entire operation time $T_L$ of LISA, we detect
no inspirals with $S/N>(S/N)_o$.

Since the relevant statistics for supermassive black holes and compact objects are poorly
known, we will not attempt a detailed calculation that allows for all possible factors
(e.g., source-orientation effects).
Instead, for a first-pass estimate of the likelihood that LISA will see a transition, we will
i) fix $M=10^6$, 
ii) approximate LISA's noise curve as flat (in other words, ignore variations in $S/N$ due to the
  emitted radiation being slightly off LISA's peak sensitivity),
iii) ignore any orientation-related increase in the emissivity of the source or the
  sensitivity of LISA,
iv) approximate $N_c$ as independent of $m$, 
v) further replace $N_c$ at each $a$ by some characteristic number of cycles (the precise value
to be chosen later, when we understand how $N_c$ varies), 
and
vi) assume all black holes have the same value of $a$ (again, to be chosen later).
To be particularly explicit, we assume the $S/N$ varies  with $m$, $d$, and $a$ in the
following manner:
\begin{eqnarray}
\label{eq:detectformulaScale}
\left(\frac{S}{N}\right)(m,d,a) &\approx& \frac{m}{10 M_\odot} \frac{1\text{Gpc}}{d}
     \left(\frac{S}{N}\right)_{\text{A}} \nonumber \\
 & =& K \frac{m}{d}  \; .
\end{eqnarray}
Here $(S/N)_A\equiv (S/N)_A(10M_\odot, 1\text{Gpc}, a)$ is a fiducial approximation to the
signal to noise ratio for an inspiral with $m=10M_\odot$, $d=1\text{Gpc}$, and $a$.

Suppose we have a discrete family of possible compact objects of masses $m_k$
with rates (per galaxy containing a $10^6 M_\odot$ hole) $r_k$;
suppose the number density of galaxies
containing a $10^6 M_\odot$ hole is $\rho_g$. Subdividing the universe into
cubes of cell size $\Delta r$, we find the probability a given cell has an inspiral of mass
$m_k$ into a $10^6 M_\odot$ hole at some time during the lifetime $T_L$ of LISA is $p_k = \rho_g r_k T_L \Delta r^3$.  Suppose we're concerned with a
threshold $S/N$ level $S/N=s_o$. At such a level we could see a source of mass $m_k$ out to a
distance 
$d_k = K m_k/s_o$.
If no inspirals have $S/N>s_o$, then for every cell in range, we have no inspirals of any mass
type. Therefore, the probability that no inspirals occur with $S/N<s_o$ is
\begin{eqnarray}
P(\text{no } S/N>s_o) &=& \prod_k (1-p_k)^{4\pi d_k^3/3 \Delta r^3}\nonumber \\
\label{eq:probabilityNoSee}
   &\approx & 
    \exp\left[ -\frac{4 \pi  R_\text{net} T_L}{3}\frac{K^3\left< m^3 \right>}{s_o^3}   \right]   
\end{eqnarray}
where in the last line we use $p_k\ll 1$, $R_\text{net}\equiv \sum \rho_g r_k$ (the net event
rate per unit volume for all inspirals), and
$<m^3>\equiv \sum \rho_g r_k m^3 / R_\text{net}$ (the mean cubed  mass of
inspiralling bodies, where weights are by event rate). Note that
$4 \pi K^3 \left<m^3\right>/(3 s_o^3)$ is the volume of space in which an inspiral involving a mass 
$\left<m^3\right>^{1/3}$ can be seen with a signal-to-noise ratio $> s_o$.
Necessarily, the probability that \emph{some} source
has $S/N > s_o$ is $P(\text{some } S/N>s_o)=1-P(\text{no } S/N >s_o)$

We can reorganize this expression to tell us, for a given probability $P_\text{no}$, at what $S/N$ we
will have a probability $P_\text{no}$ of having no signals of stronger strength:
\begin{eqnarray}
\left(\frac{S}{N}\right)_\text{no}(P_\text{no}) 
   &\equiv& \frac{\left< m^3 \right>^{1/3}}{10 M_\odot} 
     \frac{1\text{Gpc}}{\left( \frac{4\pi}{3} R_{net} T_L  \right)^{-1/3}} 
        \frac{1}{[\ln(1/P_\text{no})]^{1/3}}
    \nonumber \\
\label{eq:largestSignal}
 && \times \left(\frac{S}{N}\right)_{A} \; .
\end{eqnarray}

\subsection{Probability of detecting a transition during LISA's operation}
The $S/N$ threshold [Eq. (\ref{eq:largestSignal})] depends very sensitively (through
$\left< m^3 \right>^{1/3}$) on low-probability high-mass inspirals. By way of illustration, a
family of $0.6 M_\odot$ white dwarfs inspiralling with rate $R$ and a black hole family of mass $30 M_\odot$ and
rate $10^{-4}R$ contribute in similar proportions to $\left< m^3 \right>^{1/3}$.  
At present, the astrophysical community lacks sufficiently understanding of the
high-mass tail to be able to reliably compute $\left< m^3 \right>^{1/3}$.  Therefore, we will
neglect such objects and focus on the slightly-better understood problem of capture of
conventional compact objects. Doing so, we will underestimate the true $(S/N)_\text{no}$.

Even disregarding the high-mass holes, event rates for capture
\cite{InspiralRatesFreitag,InspiralRatesRees} remain very loosely determined, 
ranging from rates of $\sim 2\times 10^{-6}$/yr/galaxy  to $\sim 10^{-4}$/yr/galaxy.   
We take two cases as characteristic:
\begin{itemize}
\item \emph{Freitag} (F) Based on astrophysical discussion by Miralda-Escude and
  Gould \cite{InspiralRates3},
  Freitag allows for three species: white dwarfs ($m_\text{WD}=0.6
  M_\odot$, $r_{WD}\sim 10^{-5}$/yr); neutron stars ($m_\text{NS}=1.4 M_\odot$, 
  $r_{NS}\sim 2\times 10^{-6}$/yr); and black holes ($m_\text{BH}=7 M_\odot$, $r_{BH}\sim 10^{-6}$/yr).
  In this case, the net event rate $R_\text{net}$ is dominated by low-mass WD inspirals, but 
  black holes dominate the events seen by LISA.
  Using a LISA lifetime $T_L = 3\text{ yr}$ and (based on Sigurdsson and Rees's estimate that
  the density of
$10^{6}M_\odot$ holes at their cores is  around the density of spirals, since spirals have low
mass and ellipticals high mass supermassive holes \cite{InspiralRatesRees})
 $\rho_g \sim 0.003/\text{Mpc}^3$,
  we find
  \begin{eqnarray*}
  \left(\frac{S}{N}\right)_\text{no,F} &=& 
     2.67 \left(\frac{\ln(2)}{\ln (1/P_\text{no})}\right)^{1/3}
      \nonumber \\
    \label{eq:largestSignalNumberF}
      &&     \times \left(\frac{S}{N}\right)_{A}\left( 10 M_\odot, 1 \text{Gpc}\right).
  \end{eqnarray*}
\item \emph{Sigurdsson and Rees} (SR) They consider two types of galaxies --- spirals and dwarf
  ellipticals --- but only the latter leads to significant event rates.   In that case, they
  uses the following masses and rates for the three species: 
  WD ($m_\text{WD}=0.6 M_\odot$, $r_{WD}\sim 3\times 10^{-8}$/yr); 
  NS ($m_\text{NS}=1.4 M_\odot$, $r_{NS}\sim 10^{-7}$/yr); 
  and BH ($m_\text{BH}=5 M_\odot$, $r_{BH}\sim 10^{-6}$/yr).  (For black holes, these authors provide only
  an off-the-cuff estimate; we have taken some liberty in interpreting it, choosing a mildly
  optimistic characteristic black hole mass.)
  Again using $T_L = 3\text{ yr}$ and $\rho_g = 0.003/\text{Mpc}^3$, 
  we find
  \begin{eqnarray*}
  \left(\frac{S}{N}\right)_\text{no,SR} &=& 
    1.90 \left(\frac{\ln(2)}{\ln (1/P_\text{no})}\right)^{1/3}
      \nonumber \\
  \label{eq:largestSignalNumberSR}
     && \times \left(\frac{S}{N}\right)_{A}\left( 10 M_\odot, 1 \text{Gpc}\right).
  \end{eqnarray*}
\end{itemize}
In performing both calculations,  we use  Sigurdsson and Rees's estimate that the density of galaxies
with (spirals
have low-mass holes; ellipticals and others tend to have more), so  $\rho_g\sim
0.003/\text{Mpc}^3$. Also, we use a LISA lifetime $T_L = 3$ yr.

In sum, we suspect that on astrophysical grounds we will have a 50\% chance of seeing no
source with $S/N$ roughly greater than 
\begin{equation}
\label{eq:signalCutoff}
  \left(\frac{S}{N}\right)_\text{no,guess} \approx 
    2.5
 \times \left(\frac{S}{N}\right)_{A} \; ,
\end{equation}
where the $(S/N)_\text{A}$ will chosen to be the most reasonable $S/N$ over all orbital
parameters ($e$) and black hole spins ($a$), given the fiducial parameters $d=1\text{Gpc}$,
$m=10 M_\odot$, and $M=10^{6} M_\odot$.

\section{\label{sec:sch}Schwarzchild Supermassive black hole (SMBH)}
To illustrate this scheme in a case where all terms are algebraically tractable, we discuss the
range of probable transition durations when the capturing hole has no angular momentum (Schwarzchild).

\subsection{Choosing parameters}

Rather than using $E,L$ to characterize the orbit, when the orbit is confined in radius between
two turning points (i.e. when it is bound),  it is far simpler to characterize the
potential $V =-(dr/d\tau)^{2}$ by the location of its 3 roots, $r_{\pm
},\bar{r}$, where $r_{\pm }$ are the inner and outer turning points of the bound orbit and
$\bar{r}$ is the innermost root:
\begin{eqnarray}
\label{eq:schpotential}
\left( \frac{dr}{d\tau }\right) ^{2} &=&-V=\frac{1-E^{2}}{r^{3}}\left( r_{+}-r\right) \left(
r-r_{-}\right) \left( r-\bar{r}\right) \nonumber \\ &=&E^{2}-\left( 1-\frac{2}{r}\right) \left(
1+\frac{L^{2}}{r^{2}}\right)
\end{eqnarray}
Since we have only two free parameters, the three roots are not independent; they satisfy a
self-consistency polynomial.  For this reason, we introduce $p,e$ --- parameters analogous to
semi-latus rectum and eccentricity in classical mechanics. Employing a
consistency relation [generally Eq. (\ref{eq:genconstraint}) of Appendix \ref{ap:kerrbasics}]
 to
set $\bar{r}$, we find
\begin{equation}
\label{eq:schRs}
r_{\pm }\equiv \frac{p}{1\mp e}\/,\qquad \bar{r}=\frac{2p}{p-4}\/,
\end{equation}
\begin{equation}
E^{2}=\frac{\left( p-2-2e\right) \left( p-2+2e\right) }{p\left(
p-3-e^{2}\right) }
\/,\qquad L^{2}=\frac{p^{2}M^{2}}{p-3-e^{2}}.
\label{eq:schEL}
\end{equation}
These $p,e$ parameters have the notable advantage that bound orbits (orbits that cannot
escape to infinity) and
non-plunging orbits (orbits which avoid the central black hole) 
are easy to describe: bound orbits have
$e\in [0,1]$, while non-plunging orbits  have  $0<r_{-}-\bar{r}=p(p-6-2e)/[(1+e)(p-4)]$, or
\begin{equation}
\label{eq:zsch}
z=p-6-2e > 0.
\end{equation}
As one approaches the transition, the maximum of the potential $V$ decreases, $r_-$ approaches $\bar{r}$,  and  $z$
approaches $0$.  
We can
equivalently specify the location $r$ of a transition by only one of $p$ or $e$, with the other
determined by $z=0$. I will typically use $e$. For example, a transition of eccentricity $e$
occurs at radius $r=r_{-}=\bar{r}=2(3+e)/(1+e)$.

The parameters $p$, $e$ used here are identical to those used the Teukolsky-equation-based
inspiral literature \cite{CKP,Dan}. For example, the above discussion mirrors that in Cutler,
Kennefick, and Poisson \cite{CKP} between their equations (2.4) and (2.8), with the change of
notation 
$r_1\rightarrow r_-$, $r_2\rightarrow r_+$ and $r_3\rightarrow \bar{r}$.

\subsection{Dependence of  Transition Parameters on Eccentricity}
We know the potential [Eq. (\ref{eq:schpotential})]; hence we find that when 
$r_- = \bar{r}$ (=the transition radius) we have
\begin{equation}
\label{eq:vddotGeneral}
V'' = - (1-E^2)2\frac{r_+ - r_-}{r_{-}^3}
\end{equation}
and therefore, substituting $r_\pm$ and $E$ from Eqs. (\ref{eq:schRs}),
(\ref{eq:schEL}) into $\tau_o=(V''/2)^{-1/2}$ we find
\begin{equation}
\label{eq:schTau0}
\tau_o = (3+e)\sqrt\frac{2(9-e^2)}{e (1+e)^3}.
\end{equation}
Similarly, substituting  $r=r_-$ into $\gamma = -g^{tt} E$ gives us the characteristic
time required to make the transition: 
\begin{equation}
\label{eq:schGammaTau0}
\gamma \tau_o = \frac{2(3+e)^2}{\sqrt{e(1+e)^3}}
\end{equation}

Since $d\phi/dt = g^{\phi\phi}L/(-g^{tt}E) = L/r^2(1-2/r)E$, we find
\begin{eqnarray}
\left(\frac{d}{dr}\frac{d\phi}{dt}\right)_{r=r_\text{max}} &=& 
     \frac{2(r_\text{max}-3)}{r_\text{max}^{5/2}(r_\text{max}-2)}
     \; ,
\end{eqnarray}
when $E$, $L$ are consistent with a circular orbit at $r=r_\text{max}$.
Using the above and 
Eq. (\ref{eq:drMonochromatic}), we conclude that
for moderate eccentricity the natural ``transition extent'' $\delta r_\text{ref}$ is
\begin{equation}
\delta r_\text{ref} \approx \pi \frac{\sqrt{2 e (3+e)}}{(1+e)(3-e)}.
\end{equation}
This scale naturally varies with the scale of the potential
(namely, $\delta r_\text{ref}\propto 1/\sqrt{V''}$) as $e\rightarrow 0$.

Further straightforward computations using the potential [Eq. (\ref{eq:schpotential})]  reveal
how $r_\text{max}$ varies due to loss of $E$ and $L$ via wave emission when the particle is
nearly in a circular orbit (so $dE\approx \Omega dL$):
\[
\frac{d r_\text{max}}{d t} =  \frac{d L}{d t}2
\frac{(r_\text{max}-3)^{3/2}}{r_\text{max}-6} \; .
\]
We therefore can express $I_\text{ad,min}$ in terms of (tabulated) known radiation-reaction
angular momentum fluxes $\left<dL/dt\right>$:
\begin{equation}
\label{eq:schIadmin}
I_\text{ad,min} = \left( \frac{dL}{dt}\right)^2 \frac{(3-e)^2(3+e)}{2 e^2 (1+e)} \; .
\end{equation}
To obtain a rough approximation of $I_\text{ad,min}$, rather than use the true $dL/dt$
appropriate to circular orbits,  we approximate $dL/dt$ by the Peters-Mathews expression 
[Eq. (\ref{eq:dLdtPeters})].

Our scheme ceases to apply when
the eccentricity is below $e_\text{ad}$ defined by $\Delta I=I_\text{max}(e_\text{ad})$ [Eq. (\ref{eq:defineEad})]. In the special
case of $a=0$, where $p_s = 6+2e$, the definition of $I_\text{max}$
[Eq. (\ref{eq:generalImax})] reduces to 
\begin{equation}
I_\text{max} = \frac{32}{27}\frac{e^3}{(9-e^2)(1+e)} \; .
\end{equation}

\subsection{Transition duration}
With all the necessary elements assembled, we can apply our program 
[Eqs. (\ref{eq:mainGeneral}), (\ref{eq:mainGeneralCutoffs}), (\ref{eq:defineTad})] to estimate the
distribution in number of orbital cycles $N_c \equiv T_{c} \Omega / 2 \pi$ we expect when a
particle spirals into a nonspinning black hole at some fixed, known eccentricity $e$.

\begin{figure}
\includegraphics{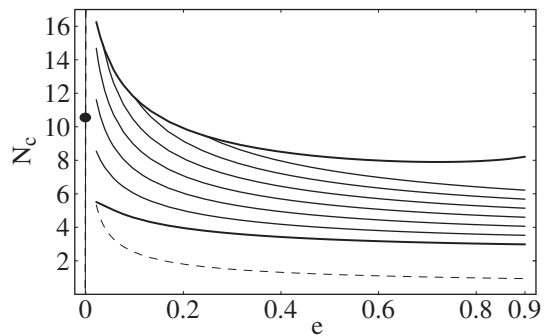}
\caption{\label{fig:schanswers} Plots of various predictions for the expected number of angular
  cycles ($N_c$) 
versus eccentricity ($e$)
for a transition from inspiral to plunge with \protect{$\eta=10^{-5}$}. 
The top solid curve is the number of cycles when \protect{$T_c=T_{c,\text{ad}}$}
[Eq. (\protect\ref{eq:defineTad})], an estimate 
of the longest possible (adiabatic) transition duration. The bottom solid curve is the number
of cycles when $T_c = T_{c,-}$ [Eq. (\protect\ref{eq:mainGeneral})].  The 6 curves in between
are the number of cycles when $T_c=T_{c,1},\ldots T_{c,6}$
[Eq. (\protect\ref{eq:mainGeneralCutoffs})]; as only a fraction $\approx 10^{-1},\ldots,
10^{-6}$ of inspirals can
have durations above these curves (respectively), they illustrate how few particles have durations
significantly differing from  $T_{c,-}$.  
The dot indicates the Ori-Thorne (OT) prediction for circular equatorial inspiral.
The dashed curve is
a   characteristic-scale-based prediction based on \protect{$T_c=4\gamma \tau_o$}, used to illustrate the
  significance of the ``logarithmic correction.''
The plot starts at \protect{$e=e_\text{ad}\approx 0.0215$}, at which point $\Delta I=I_\text{max}$.  
}
\end{figure}

The results for $\eta =10^{-5}$ are shown
in Fig. \ref{fig:schanswers}.
When our adiabatic approximation applies, we find that to a good approximation (within around 1
cycle) most transitions should have duration close to the shortest transition duration $N_{c,\text{ad}}=T_{c,\text{ad}} \Omega
/2 \pi\in[3,5]$.  
In particular, within the region $e>e_\text{ad}$ where our adiabatic approximation applies,
almost all transitions will last for less than the Ori-Thorne (OT) circular duration; most will
last for substiantially less.
At low eccentricity, most transitions seem to approach a result somewhat different than the OT
circular estimate. Since OT use a different convention for $\delta r$
\footnote{Since all results depend (mildly) on the convention for transition extent, and since 
the Ori-Thorne prediction implicitly employs a characteristic length
$\delta r_\text{ref}\approx$ (few) $\times R_o \propto \eta^{2/5}$
with $R_o$ given by Ori-Thorne Eq. (3.20), while the ``standard'' predictions
[Eqs.
(\ref{eq:mainGeneral}),(\ref{eq:mainGeneralCutoffs})]
  use Eq. (\ref{eq:drMonochromatic}), with $\delta r_\text{ref} \propto \eta^0$, we cannot guarantee that
 the results should be precisely compatible.
}
and since significant
changes could still occur in the fundamentally nonadiabatic region between $e=0$ and
$e=e_\text{ad}$,
we do not find the discrepancy troubling.

In the above, we show results for only $\eta=10^{-5}$ (say, for  $m=10M_\odot$ and $M=10^{6}$). As the variation of
the duration with $\eta$ is weak ---  we find
$H(\eta_o=10^{-5})\in [0.1,0.4]$) [Eq. (\ref{eq:mainVaryWithEta})] --- even substantially
different test particle masses (e.g., $m\in[0.1,30]M_\odot$ with $M=10^{6}$) lead to results of
the form above, scaled up or down by a factor $\lesssim 2$.

\subsection{Prospects for LISA detecting a given transition}
As discussed in Sec. \ref{sec:framework2}, we can estimate the signal-to-noise ratio for a
given transition using Eq. (\ref{eq:detectformula}).  For the standard case of a $10 M_\odot$
particle falling into a $M=10^6 M_\odot$,
application of that formula reveals no 
higher $S/N$ than that predicted by Ori and Thorne; moreover, barring astrophysically unlikely
masses, all transitions have too low a
$S/N$ to be detected. [See Fig. \ref{fig:detect} below for details.]  For example, if 
an inspiral of mass $m$ into a $10^{6}M_\odot$ hole occurs at the fiducial distance $1$ Gpc
with $e=1/3$, we have a 90\% chance that $S/N\in [0.91,1.01] (m/10 M_\odot)$.

The results for $S/N$ can be well-approximated by way of Eq. (\ref{eq:detectformulaRelative}) and
a comparison with Ori and Thorne's
results for circular inspiral. (See Appendix \ref{ap:OTreference} for a summary of OT results).
Specifically, using the fiducial case of
$10M_\odot$ on $10^{6}M_\odot$ at $1$ Gpc, for which we have  $(S/N)_{OT}=1.6$ and
$N_{c,OT}=10.5$, we find the general $S/N$ for captures by a $M=10^{6} M_\odot$ hole
to be about
\begin{eqnarray*}
(S/N)&\approx& 1.6 \times \sqrt{\frac{N_{c}}{10.5}} \times
\frac{m}{10M_\odot}  \frac{1\text{Gpc}}{d} \; .
\end{eqnarray*}

\section{\label{sec:kerr}Kerr SMBH}
The Kerr case follows similarly, 
save with an additional parameter ($a$).

\subsection{Parameterizing Orbits, Potential}

As before, it is simplest to characterize the potential by its three roots $r_{\pm}=p/(1\mp
e)$, $\bar{r}$:
\begin{equation}
\left( \frac{dr}{d\tau }\right) ^{2}=-V=\frac{1-E^{2}}{r^{3}}\left( r_{+}-r\right) \left(
r-r_{-}\right) \left( r-\bar{r}\right).
\end{equation}
And as before we can define $r_{\pm}=p/(1\mp e)$; as before, we find a self-consistency
relation $P(p,e,a,\bar{r})$ [Eq. (\ref{eq:genconstraint})], permitting us to solve for
$\bar{r}(p,e,a)$.
As before, we can characterize the proximity to the last-stable surface by way of the
separation between the two innermost roots ($r_- - \bar{r}$).  
Finally, as before, for each fixed black hole ($a=$const) and
each particle exactly on the transition line from orbit to plunge, the particle can have $e\in
[0,1)$; its $p$ will be constrained by the analogue of the Schwarzchild $p=6+2e$: the
self-consistency relation Eq. (\ref{eq:constraint}), which implicitly defines $p_s(e,a)$ such that
$\bar{r}(p_s,e,a)=p_s/(1+e)$.

\subsection{Dependence of transition parameters on $e$,$p$}
Since the potential has the same structure as before, the same general expression
Eq. (\ref{eq:vddotGeneral}) applies, with $E$ now determined by expressions in Appendix
\ref{ap:kerrbasics}.  By explicitly differentiating the potential [Eq. (\ref{eq:potential})],
inserting the definitions $r_{\pm}=p/(1\mp e)$, and demanding the inner turning point is a
maximum (so $\bar{r}=r_- = p/(1+e)$), we find
\begin{equation}
V'' = -8e \frac{(1+e)^3}{(3-e)p_s^3}
\end{equation}
and therefore know $\tau_o=(V''/2)^{-1/2}$ in terms of $p,e$ at the transition.

The $\gamma$ factor follows from inserting $r=p/(1+e)$ into the usual expression for
$dt/d\tau$:
\begin{equation}
\label{eq:kerrDtDtau}
\gamma\equiv\frac{dt}{d\tau} = - g^{tt} E + g^{t\phi} L
\end{equation}
where $g^{tt}$ and $g^{t\phi}$ are known Kerr metric coefficients.
Here, $E$ and $L$ are evaluated using the expressions (\ref{eq:genEnergy}) and
(\ref{eq:genL})
 discussed in Appendix
\ref{ap:kerrbasics}, with $\bar{r}=p_s/(1+e)$.

We obtain the transition extent $\delta r_\text{ref}$ with the usual Eq. (\ref{eq:drMonochromatic}).
This requires $\gamma$ [Eq. (\ref{eq:kerrDtDtau}), above], $\tau_o$ (also above), and
$d(d\phi/dt)/dr$ [Eq. (\ref{eq:instantaneousFrequency})].

Finally, as in the Schwarzchild case we estimate $I_\text{ad,min}$ [Eq. 
(\ref{eq:defineIadmin})] and thus $T_{c,\text{ad}}$ [Eq. (\ref{eq:defineTad})]
 via i) expressing $d r_\text{max}/d\tau$ in terms of  $dL/dt$ using explicit expressions
 for $r_\text{max}(E,L)$ and $dE=\Omega dL$, giving 
\begin{equation}
\frac{d r_\text{max}}{dt} =   \frac{dL}{dt} 
  \left.
  \frac{2 r (2 a \sqrt{r} + r^2-3r)^{3/2} }{(r^{3/2}+a)(r^2 -6r + 8 a \sqrt{r} - 3 a^2)}
  \right._{r=r_\text{max}}
\; ,
\end{equation}
(where we have used the orbital parameters $E$,$L$ consistent with a
circular orbit at $r=r_\text{max}$ \cite{CircularOrbits});
then ii) using the
 Peters-Mathews expressions for $dL/dt$ [Eq. (\ref{eq:dLdtPeters})] to construct an approximate
 expression for $d r_\text{max}/dt$, which we then iii) insert in Eq. (\ref{eq:defineIadmin})
 to estimate the boundary between adiabatic and nonadiabatic transitions.

\subsection{Transition duration}
\begin{figure}
\includegraphics{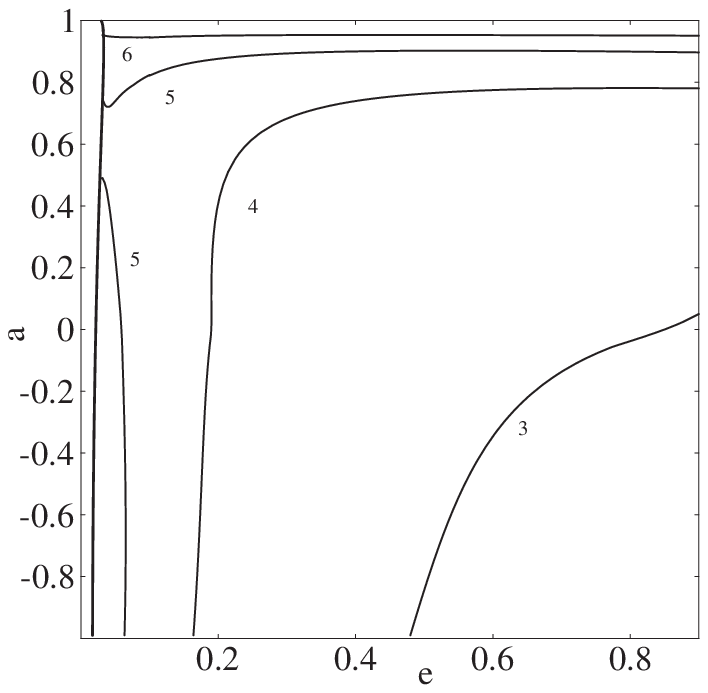}
\includegraphics{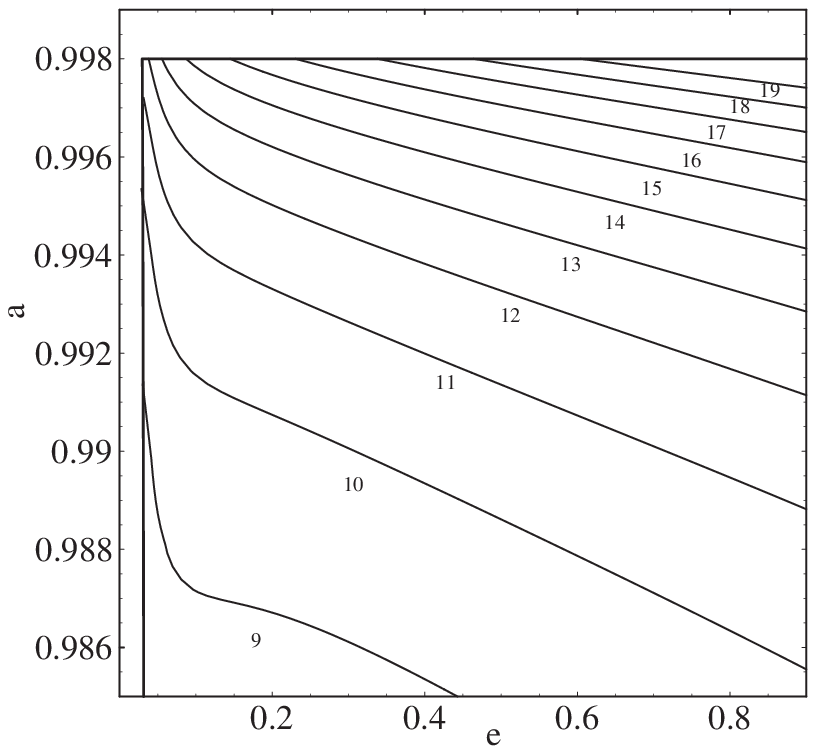}
\caption{\label{fig:kerranswers}
These two plots illustrate the shortest possible number of cycles $N_c=\Omega T_{c-}/2\pi$ 
[Eq. (\ref{eq:mainGeneral})] a
transition could last, versus eccentricity
($e$) and black hole angular momentum $(a)$ for the fiducial source 
($10 M_\odot$ into $10^{6}M_\odot$).
In both plots, contours are cut off, and  bounding curves  appear (shown heavy solid), when 
$\Delta I = I_\text{ad,min}$ and when $a=0.998$.
}
\end{figure}

\begin{figure}
\includegraphics{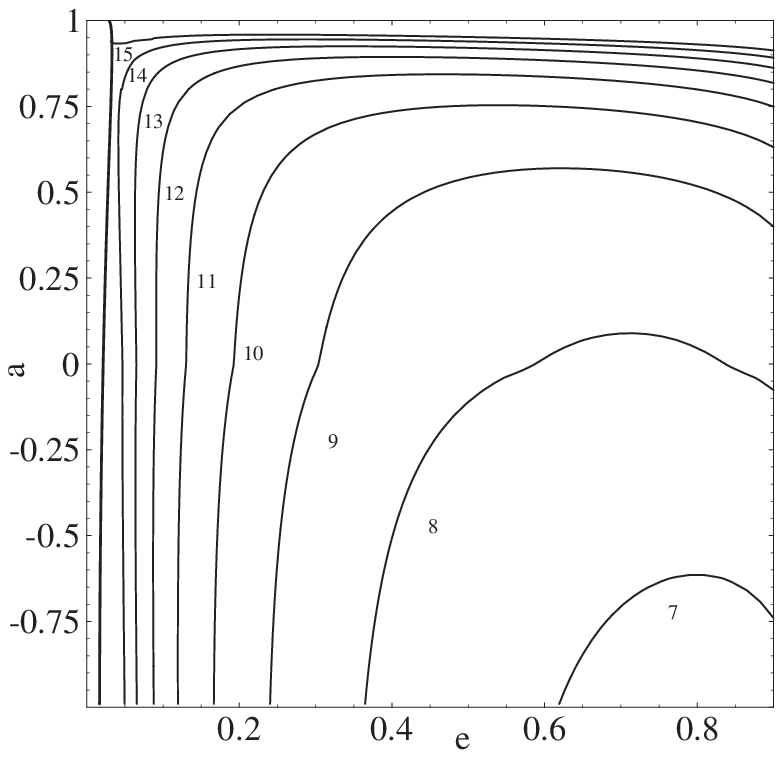}
\includegraphics{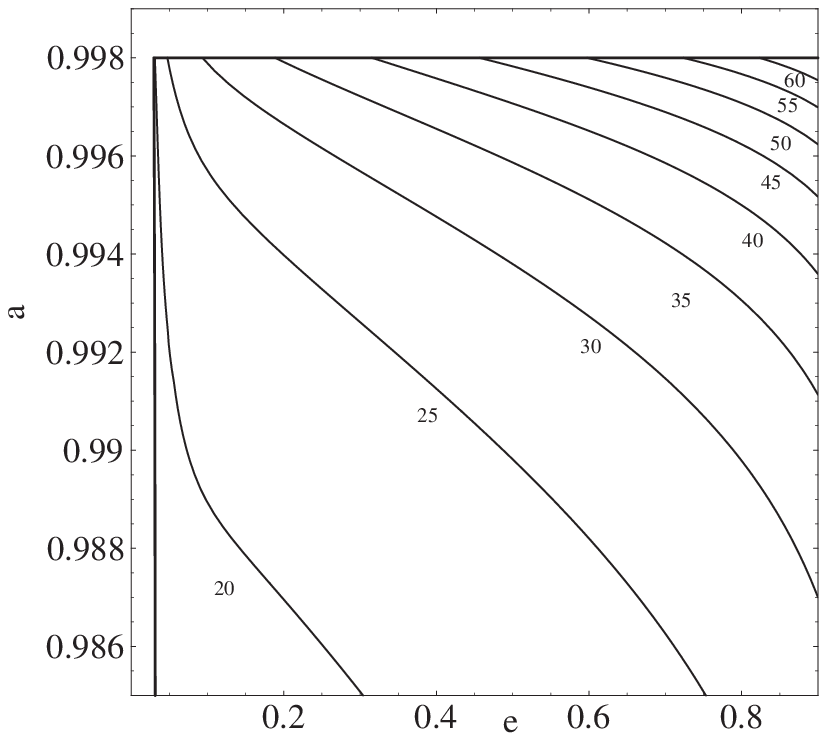}
\caption{\label{fig:kerranswers2}
These two plots illustrate the longest possible number of cycles an (adiabatic) transition
could last $N_c=\Omega T_{c,\text{ad}}/2\pi$ [Eq. (\ref{eq:defineTad})], versus eccentricity
($e$) and black hole angular  momentum ($a$)
for the fiducial source 
($10 M_\odot$ into $10^{6}M_\odot$).  Transitions of such long duration are extremely unlikely
unless $e\approx e_\text{ad}$ [see Eq. (\ref{eq:defineEad}), (\ref{eq:mainGeneralDistribution})].
In both plots, contours are cut off, and  bounding curves  appear (shown heavy solid), when 
$\Delta I = I_\text{ad,min}$ and when $a=0.998$.
}
\end{figure}

\begin{figure}
\includegraphics{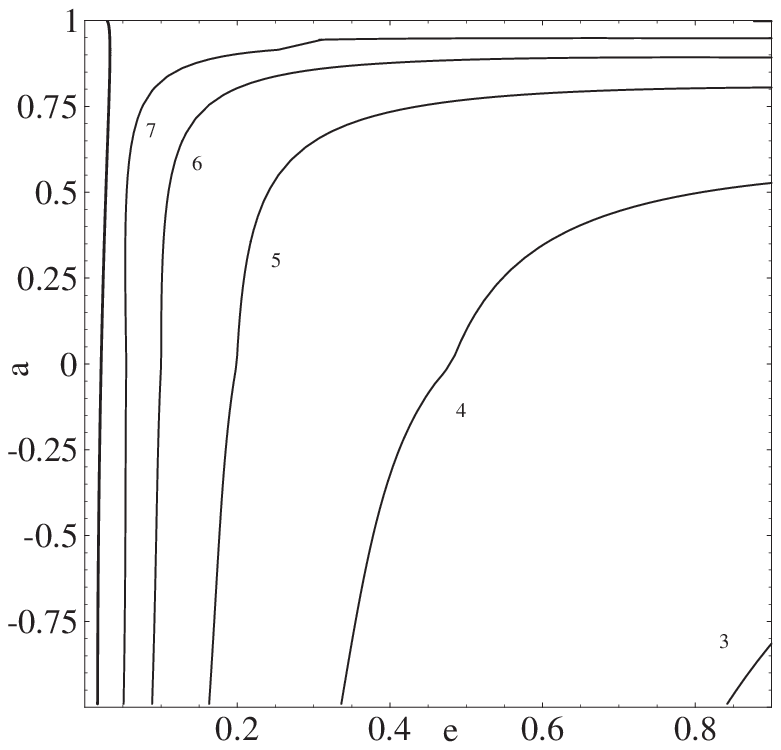}
\includegraphics{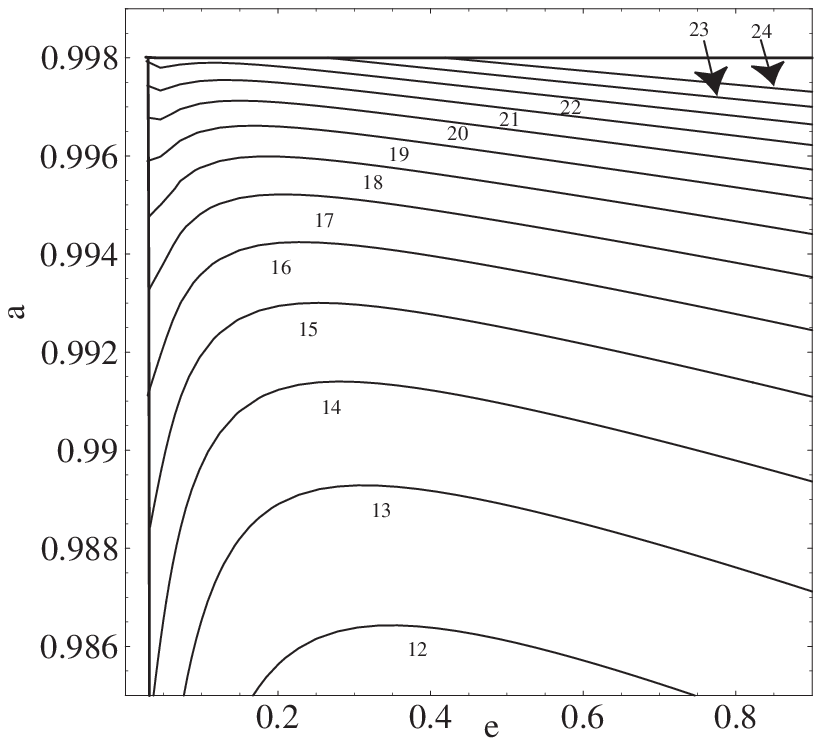}
\caption{\label{fig:kerranswers3}
These two plots illustrate $N_c =\Omega T_{c,1}/2\pi$ versus $e$ and $a$ for the fiducial case.
When the results of Fig. \ref{fig:kerranswers2} are appreciably larger than these  --- that is,
everywhere except near the left boundary $e=e_\text{ad}$ --- $\sim$ 90\%
of transitions with orbital parameters $e$, $a$ will take less than the number of cycles shown
in this plot to traverse the transition region.
}
\end{figure}

Combining these together, we can deduce the range of plausible transition durations for a test
particle of eccentricity $e$ falling into a hole of angular momentum $a$, measured
 as number of orbital cycles
$N_c(e,a)=T_c \Omega/2\pi$.
Plots of the number of cycles appropriate to $T_c=T_{c-}$ [Eq. (\ref{eq:mainGeneral})], to 
$T_c = T_{c,\text{ad}}$ [Eq. (\ref{eq:defineTad})], and to
$T_c=T_{c,1}$ [Eq. (\ref{eq:mainGeneralCutoffs})]
appear in Figs. \ref{fig:kerranswers}, \ref{fig:kerranswers2}, and \ref{fig:kerranswers3},
respectively. 

These plots all assume a fiducial source ($10 M_\odot$ on $10^{6} M_\odot$).
In these plots, we truncate the range of $e$, $a$ allowed because, 
i) we need $e$ larger than
$e_\text{ad}$ [Eq. (\ref{eq:defineEad})]; and
ii) realistic
astrophysical black holes have $a\le 0.998$ \cite{LargestA}.  Also, in these plots, we do not
extend to $e\approx 1$ because we do not have data for $I$ in this region, nor do we expect our
estimate of $\Delta I$ [Fig. \ref{fig:deltaIcurve}] to be reliable in this extreme. 

At each $a$, we see behavior largely similar to the Schwarzchild results 
discussed in Sec. \ref{sec:sch}:
i) almost all transitions take less time than the Ori-Thorne result for $e=0$; ii) as we increase the eccentricity, the
transition duration decreases; and iii) since $T_{c,1}\approx T_{c,-}$ (compare
Figs. \ref{fig:kerranswers} and \ref{fig:kerranswers3}), most transitions last close to the
shortest-possible transition duration.

\subsection{\label{sec:detectFixedSource}Prospects for LISA detecting a given transition}
\begin{figure}
\includegraphics{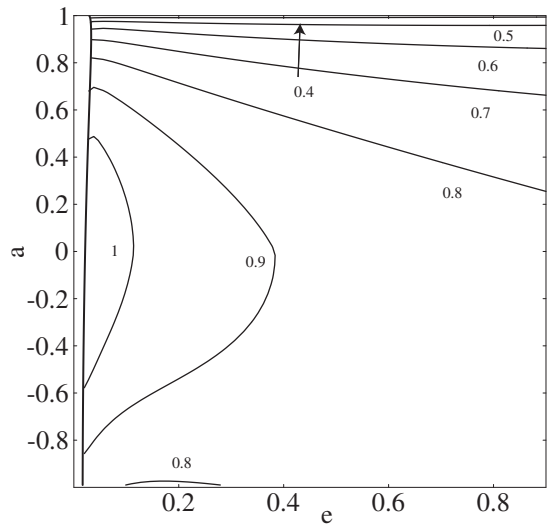}
\caption{\label{fig:detect}
This plot illustrates   $S/N$ [Eq. (\protect\ref{eq:detectformula})], given  a fiducial
source ($10 M_\odot$ into $10^{6} M_\odot$, at 
$1\text{Gpc}$) given that the black hole has angular momentum $a$, the transition occurs at
eccentricity $e$, and given the transition takes the shortest possible time (or $T_c=T_{c,-}$)
[Eq. (\ref{eq:mainGeneral})]. 
 As argued in the text, this time (and thus this $S/N$)
 will be to a good approximation
characteristic of all transitions with those interaction parameters.
}
\end{figure}

As in the Schwarzchild case, since eccentric usually transitions last for
fewer angular cycles than their circular analogues, they are less detectable
as well.  Thus, in the fiducial case of captures of a $10 M_\odot$ hole by a $10^{6} M_\odot$
hole, the data from Ori-Thorne Table II  provides an upper bound on the 
S/N seen by LISA (shown in Fig. \ref{fig:detect}).  Since this bound is small, we have little chance of seeing any given
transition.

One should notice, however, that the distribution of $S/N$ with orbital parameters is very
flat and not much below $1$. Therefore, only a modest improvement in LISA's noise spectrum
$S_h$ could render most  (measured by volume of parameter space) of the transitions detectable.

\subsection{\label{sec:detectSomething}On probability of detection}
Because LISA at present has so poor prospects for detecting the ``fiducial'' source ($m=10
M_\odot$ at $1$Gpc), it has a poor chance
of seeing any source at all.  Even assuming all LISA sources had orbital parameters chosen to
give the longest-plausible transition length (the OT circular inspiral duration, which has
$S/N_A \lesssim 1.6$),
by the estimate of Eq. (\ref{eq:signalCutoff}) we expect we have
a $\sim$50\% chance of  no signal with $S/N \gtrsim 4$ being present in the datastream. 
With more realistic orbital parameters, we would expect a $\sim$50\% chance of no signal
$\gtrsim 2.3$.
In other words, LISA has a good to excellent chance of \emph{not} seeing any transitions from
inspiral to plunge in its lifetime.

A modest improvement in LISA's noise curve, however, would make a few circular (and to a lesser
degree eccentric) transitions from inspiral to plunge detectable.

\section{Summary}
This paper has introduced a framework (depending on observational or other conventions)
that extends the Ori-Thorne prediction for the transition
duration from inspiral to plunge to include eccentric orbits. 
While the framework and applications
contain many oversimplifications --- most notably, the fit to $\Delta I(e,a)$ and  
and the lack of a physically meaningful
convention for $\delta r_\text{ref}$ 
--- the essential physics should be captured by Sec. \ref{sec:framework}.

This paper then applies
that framework  to probable LISA sources to  suggest
that, because an eccentric transition is generally only slightly briefer than a circular one, LISA
should have  only slightly worse prospects to resolve the transition from 
inspiral to plunge for  eccentric orbits than for circular ones.  
While the prospects for detecting circular (and hence eccentric) transitions with LISA
are not good, they are not necessarily bad: modest changes to the LISA noise floor could render
a signal marginally detectable. Therefore, 
 more detailed investigations could be of use.

Potentially, we could use other portions of orbits that pass close to the hole --- for
example, the previous few ``bounces'' off the inner portion of the radial potential --- as 
probes of the strong-field metric.  Analyzed separately
(using the same $\Delta I$ framework)
 each of these 
``bounces'' should provide in itself at best of order the same S/N as the transition.
If the source has already been detected with good confidence, we should be able to coherently
integrate over many such bounces and build up excellent $S/N$.

Finally, we could hope that eccentric inclined orbits might, by some happenstance of
parameters, admit a regime of significantly longer transition times. The prospect seems
unlikely, but the author may address it in a future paper.

\appendix
\section{\label{ap:evolveMax}Evolution of the maximum}
As the conserved constants $E$, $L$ evolve, the height of any local maximum in the potential
$V$ [Eq. (\ref{eq:geodesic})] will similarly evolve. Because we are at a local maximum, we can
find a simple expression for the rate of change of the value of the potential at that local maximum:
\begin{equation}
V_{\max}(E,L) = V(r_{\max}(E,L),E,L)
\end{equation}
(where $r_{\max}$, the location of the potential's local maximum, is a solution to $dV/dr=0$). 
In general, to order of magnitude, we expect it goes as $\sim V/\tau_\text{gw}$ for
$\tau_\text{gw}$ the gravitational wave timescale $\tau_\text{gw}\sim E/\dot{E}\sim
L/\dot{L}$.  But when the particle is nearly on a circular orbit, then the source of radiation nearly
satisfies helical symmetry and therefore $dE\approx \Omega dL$ for $\Omega$ the angular
frequency of the circular orbit.  And in these special conditions $V$ changes even more slowly
than we would normally expect.

To be explicit, we evaluate $dV_{\max}/dt$, which (because we are at a maximum) we can
generally express as follows:
\begin{equation}
\label{eq:changeInVmax}
\frac{dV_{\max}}{dt} 
  = \frac{\partial V}{\partial L} 
   \left[\frac{dL}{dt} + \frac{dE}{dt}\frac{\partial V/\partial E}{\partial V/\partial L}
  \right]_{r=r_{\max}} \; .
\end{equation}

We can most transparently prove the necessary result  by rewriting the potential $V$ more
abstractly than the standard form of Eq. (\ref{eq:potential}).
Recall one derives the radial potential from the constancy of the test-particle's rest mass 
[e.g., $g^{ab}p_a p_b=-m^2$].  Since the Kerr metric in  boyer-lindquist coordinates
has form $g_{ab}=g_{tt} dt^2 + 2 g_{t\phi} dt d\phi + g_{\phi\phi}d\phi^2 + g_{rr}dr^2
+g_{\theta\theta} d\theta^2$, by employing the definitions of $E$ and $L$ (e.g., $-p_t\equiv$
energy)  and the existence of an equatorial orbit, one obtains the (first integral of the)
radial geodesic equation Eq. (\ref{eq:geodesic}) with
$V=(1+E^2 g^{tt} + L^2 g^{\phi\phi}- 2EL g^{t\phi})/g_{rr}$.
One can similarly show, by employing the definitions of the ``raised'' components (e.g., $p^t
\equiv m dt/d\tau$ for $\tau$ proper time), that
$d\phi/dt=(g^{\phi\phi}L-g^{t\phi}E)/(-g^{tt}E+g^{t\phi}L)$ and $dt/d\tau = - g_{rr} \partial
V/\partial E$.
Using these expressions in  Eq. (\ref{eq:changeInVmax}), we conclude that
\begin{equation}
\frac{dV_{\max}}{dt} = - \frac{dt/d\tau}{g_{rr}}
   \left [ 
     \frac{dL}{dt} -  \frac{dE}{dt} \frac{1}{d\phi/dt}
   \right ]_{r=r_{\max}}.
\end{equation}
When the potential admits a nearly-circular orbit (angular frequency $\Omega=d\phi/dt$) near
the local maximum, we have  $dE\approx \Omega dL$ in the emitted radiation and therefore
$dV_\text{max}/dt$ is smaller than normal.
[Similar arguments apply to the minimum, and prove that stable circular orbits evolve to stable circular orbits.]

\section{\label{ap:kerrbasics}Kerr Parameters and Constraints}
In the Teukolsky-equation-based inspiral literature \cite{CKP,Dan}, when the orbit is
 \emph{bound} (=does not fall into hole or escape to infinity) it is characterized not by 
physical parameters ($E$,$L$,$a$) but by the location of its radial turning points ($r_\pm$)
 and the remaining root of its potential ($\bar{r}$):
\begin{eqnarray}
\label{eq:apPotential}
V&=&\frac{E^{2}-1}{r^{3}}\left(
r_{+}-r\right) \left( r-r_{-}\right) \left( r-\bar{r}\right) \\
&=&-\left( E^{2}-1\right) -\frac{2}{r}
+\frac{\left[L^{2}-a^{2}\left( E^{2}-1\right) \right]}{r^{2}} \nonumber \\
& &-\frac{2\left( L-aE\right) ^{2}}{r^{3}} 
\end{eqnarray}
To further simplify the algebra involved, one replaces $r_{\pm}$ by a parameterization analogous
to classical mechanics (semi-latus rectum and eccentricity):
\begin{equation}
r_{\pm} = \frac{p}{1\mp e}
\end{equation}
(with $p$,$e$ both real, positive).
After replacing $r_{\pm }$ by $p,e$ using $r_{\pm }=p/(1 \mp e)$, we find
the following explicit forms for $E,L$ in general: (specify retrograde
orbits by a negative $a$ sign)
\begin{eqnarray}
\label{eq:genEnergy}
E  &=&\sqrt{1-\frac{\left( 1-e^{2}\right) }{2p+\left( 1-e^{2}\right) \bar{r}}}
 \\
\label{eq:genL}
L  &=&\left[ \text{sign}(a)\frac{p\sqrt{\bar{r}}}{\sqrt{2p+\left( 1-e^{2}\right) \bar{r}}}%
+aE\right]
\end{eqnarray}

The parameters $p,e,\bar{r}$, however, are not fully independent.
The coefficients of $V\left( r\right) $ satisfy a polynomial equation when
the coefficients are expressed in terms of $L,E,a$; the coefficients must
satisfy the same polynomial when the coefficients are written using $r_{\pm
},\bar{r}$. Re-expressing that polynomial in terms of $p,e$, we find
\begin{eqnarray}
\label{eq:genconstraint}
0&=&p^{2}\left[ p\left( \bar{r}-2\right) -4\bar{r}\right] ^{2}
    +a^{4}\left[p+\left( 1-e^{2}\right) \bar{r}\right] ^{2} \nonumber \\
& &-2a^{2}p\left\{ 4\left(1-e^{2}\right) \bar{r}^{2} 
   +2p^{2}\left( 2+\bar{r}\right) \right.\nonumber \\
& &+p\bar{r}\left. \left[
8+\left( \bar{r}-2\right) \left( 1-e^{2}\right) \right] \right\}
\end{eqnarray}
As a practical matter, we usually specify a particle by ($p$,$e$,$a$) and
then solve for $\bar{r}$ and hence all other orbital parameters.

\subsection{Separatrix}
The transition from stable to unstable occurs when $\bar{r}=r_{-}=p/(1+e)$.  
We can attempt to express this relation in terms of the usual orbital 
parameters ($p$,$e$,$a$). 
Substituting in the above polynomial [Eq. (\ref{eq:genconstraint})], we get
\begin{eqnarray}
\label{eq:constraint}
0=&p_s^{2}\left( p_s-6-2e\right) ^{2} +a^{4}\left( e-3\right) ^{2}
\left(1+e\right) ^{2}\nonumber\\
&-2a^{2}\left( 1+e\right) p_s\left[14+2e^{2}+p_s\left( 3-e\right) \right]
\end{eqnarray}
where $p=p_s(e,a)$, defined by (appropriate solutions to) the above relation, determines the
separatrix between stable and unstable geodesic orbits.

\subsection{Minimum of potential}
At the transition from stable to unstable orbits, the maximum of the potential is at
$r_-=\bar{r}=p_s/(1+e)$. Therefore, by symbolically differentiating 
  Eq. (\ref{eq:apPotential}), we find $r_{\min}=3p_s/(3-e)$.
Inserting this expression into $V$ and noting
  $V(r_{\max})=0$, we find $I_{\max}\equiv V(r_{\max})-V(r_{\min})$:
  \begin{equation}
    \label{eq:generalImax}
    I_{max}= \left( \frac{4 e}{3} \right)^3 \frac{1}{(3-e)(1+e)p_s}
  \end{equation}

\section{Results from Ori and Thorne}
\label{ap:OTreference}
So that the reader can more easily compare our estimates against those of Ori and Thorne, we provide
approximations to their results.

In Ori and Thorne's Table II, they tabulate $S/N$ and $N_c$ for $10 M_{\odot}$ on
$10^{6} M_\odot$, using fiducial distance $1$ Gpc to set the amplitude scale.  
[In this table, OT list the number of quadrupolar gravitational wave cycles $N_\text{cyc}=2 N_c$.]
We can
approximate their results by the two functions
\begin{eqnarray}
\label{eq:otApproxNc}
\log_{10} N_{c,\text{OT}} &\approx&   
   0.92 + 0.110 x + 0.0478 x^2 \;,
        \quad a\in[0.9,0.99]   \nonumber \\
   &\approx&      
   1.03 + 0.073 x + 0.138 x^2 \; ,
        a<0.9   
\end{eqnarray}
and 
\begin{eqnarray}
\label{eq:otApproxSN}
\log_{10} &\left( \frac{S}{N}\right)_\text{OT}& \times 10^{-22}  \nonumber \\
   &\approx&   
   0.3156 + 0.3193 x + 0.0451 x^2 \;,
        \quad {a\in[0.8,0.99]}   \nonumber \\
   &\approx&      
   0.191 - 0.226 x  -  0.448 x^2 \; ,
         {a<0.8}   
\end{eqnarray}
where $x\equiv \log_{10} (1-a)$.

\begin{acknowledgements}
We thank Kip Thorne for helpful discussions and advice, and Dan Kennefick for the 
kindness with which he provided invaluable numerical data throughout the development of
this paper.
\end{acknowledgements}

\end{document}